\documentclass[prd,tightenlines,floatfix,showpacs,preprintnumbers,nofootinbib,eqsecnum,superscriptaddress,preprint]{revtex4-1}

\usepackage{epsfig}
 
\newcommand{\beq}{\begin{equation}}
\newcommand{\eeq}{\end{equation}}
\newcommand{\bea}{\begin{eqnarray}}
\newcommand{\eea}{\end{eqnarray}}

\begin{document}

\preprint{MS-TP-18-19}

\title{Neutralino/chargino pair production at NLO+NLL with
 resummation-improved PDFs for LHC Run II}

\author{J.\ Fiaschi}
\email{fiaschi@uni-muenster.de}
\affiliation{Institut f\"ur Theoretische Physik, Westf\"alische
 Wilhelms-Universit\"at M\"unster, Wilhelm-Klemm-Stra{\ss}e 9,
 D-48149 M\"unster, Germany}

\author{M.\ Klasen}
\email{michael.klasen@uni-muenster.de}
\affiliation{Institut f\"ur Theoretische Physik, Westf\"alische
 Wilhelms-Universit\"at M\"unster, Wilhelm-Klemm-Stra{\ss}e 9,
 D-48149 M\"unster, Germany}

\date{\today}


\begin{abstract}
We make use of recently released parton density functions (PDFs) with
 threshold-resummation improvement to consistently calculate theoretical
 predictions for neutralino and chargino pair production at next-to-leading
 order and next-to-leading logarithmic accuracy. The updated cross sections
 have been computed for experimentally relevant higgsino and gaugino search
 channels at the ongoing Run II of the LHC. A factorisation method is applied
 to exploit the smaller PDF uncertainty of the global PDF sets and to avoid
 complications arising in the refitting of threshold-resummation improved PDF
 replicas in Mellin space. The reduction of the scale uncertainty due to the
 resummation is, however, explicitly taken into account. As expected, the
 resummation contributions in the PDF fits partially compensate the cross
 section enhancements induced by those in the partonic matrix elements.
\end{abstract}


\maketitle


\section{Introduction}
\label{sec:1}

The Minimal Supersymmetric Standard Model (MSSM) is a theoretically and
phenomenologically well motivated extension of the Standard Model (SM)
of particle physics \cite{Nilles:1983ge,Haber:1984rc}. It predicts in
particular fermionic partners of the neutral and charged gauge and Higgs
bosons called gauginos and higgsinos, whose lightest neutral mass
eigenstate, the lightest neutralino, is one of the best studied dark
matter candidates \cite{Klasen:2015uma,Fuks:2008ab,Herrmann:2009wk,%
Herrmann:2009mp}. Heavier neutralinos and charginos decay typically
into multilepton final states and missing transverse momentum.
Searches for gaugino- \cite{Aaboud:2017nhr,Sirunyan:2017nyt} or higgsino-like
particles \cite{Aaboud:2018zeb,Sirunyan:2018iwl} are important physics
goals at the LHC. They are often carried out in the framework of
simplified models \cite{Alwall:2008ag,Calibbi:2014lga}. Care must,
however, be taken that the theoretical assumptions are not overly
simplified \cite{Fuks:2017rio}.

Experimental measurements of supersymmetric (SUSY) production cross
sections at the ongoing Run II of the LHC require precise theoretical
calculations at the level of next-to-leading order (NLO) QCD and beyond
\cite{Beenakker:1996ch,Beenakker:1997ut,Beenakker:1999xh,Berger:1999mc,%
Berger:2000iu,Spira:2002rd,Jin:2003ez,Binoth:2011xi,Ahmadov:2013qga,%
Demirci:2014gva}. In the
perturbative expansion, logarithmically enhanced terms appear beyond
leading order in the strong coupling constant $\alpha_s$, whose
contributions can be sizeable close to production threshold or at small
transverse momentum of the produced SUSY particle pair. Their effect on
neutralino, chargino \cite{Li:2007ih,Debove:2009ia,Debove:2010kf,%
Debove:2011xj,Fuks:2012qx,Fuks:2016vdc}, slepton \cite{Yang:2005ts,%
Broggio:2011bd,Bozzi:2006fw,Bozzi:2007qr,Bozzi:2007tea,Fuks:2013lya,%
Bonvini:2015ira,Fiaschi:2018xdm}, squark, gluino \cite{Beenakker:2014sma,%
Borschensky:2014cia,Beneke:2016kvz,Beenakker:2016lwe}, stop
\cite{Broggio:2013cia,Beenakker:2016gmf} and also new gauge boson
production \cite{Fuks:2007gk,Jezo:2014wra,Mitra:2016kov,Klasen:2016qux}
has been taken into account to
all orders with resummation techniques to next-to-leading logarithmic
(NLL) accuracy and beyond, and the results for the electroweak production
channels have been made publicly available with the code RESUMMINO
\cite{Fuks:2013vua}.
Parton showers also partially resum logarithmically enhanced contributions,
and complementary NLO calculations with parton showers have been performed
for squark \cite{Gavin:2013kga,Gavin:2014yga}, slepton \cite {Jager:2014aua}
and gaugino pair production \cite{Baglio:2016rjx,Baglio:2017wka}.
The effect of higher order QCD corrections is
generally to enhance the theoretical estimations for the cross sections,
while on the other hand they reduce the dependence of the results on the
choice of the unphysical renormalisation and factorisation scales.

A consistent prediction of hadronic cross sections would in principle
require the same precision in the calculation of the partonic matrix
elements of the specific process as in the determination of the parton
density functions (PDFs). For this purpose some examples of
threshold-resummation improved PDFs have been recently released. In
this work we will make use of the NNPDF30\_nlo\_disdytop and
NNPDF30\_nll\_disdytop PDF sets \cite{Bonvini:2015ira}, which have been
obtained by the NNPDF collaboration with partonic matrix elements
computed respectively at NLO and NLO+NLL in the fits of experimental
data. Unfortunately, NLO+NLL accuracy is only readily available for
a rather small subset of relevant processes. Therefore, only part of
the available experimental data can enter in a consistent NLO+NLL fit.
In particular, the resummed PDF fits rely only on data from
deep-inelastic scattering (DIS), the Drell-Yan (DY) process and top-quark
pair production. Consequently, the reduction of the input data set leads
to a larger PDF error in comparison with a global PDF set.

In this work, we present updated and consistent results for the cross
sections of neutralino and chargino production at NLO+NLL for Run II
of the LHC. We adopt the $K$-factor method already discussed in the
literature \cite{Beenakker:2015rna}, which will allow for a straightforward
rescaling of the fixed-order calculations to consistent NLO+NLL results.
The neutralinos and charginos are in general a linear superposition
of SM superpartners. Following recent experimental analyses, we consider
here two specific scenarios where they are dominated by either the higgsino
or the gaugino content. Moreover, we will work with the simplified models
used in the experimental analyses of the ATLAS and CMS collaborations, where
the produced neutralinos and charginos are similar in mass, while most other
SUSY particles are decoupled. While the gauginos are assumed to decay to the
lightest SUSY particle (LSP) through intermediate (s)taus
\cite{Aaboud:2017nhr}, the higgsinos are assumed to decay to a not much
lighter LSP producing soft leptons \cite{Sirunyan:2018iwl}. The associated
experimental analyses lead to quite different constraints on the SUSY
particle masses.

This paper is organised as follows: In Sec.~\ref{sec:2}, we review the
$K$-factor method to merge the effect of a consistent resummation both in
the PDFs and in the partonic matrix elements with the smaller PDF error
of the global set and with a reliable estimate of the scale uncertainty.
In Sec.~\ref{sec:3}, we present our numerical results for differential and
total cross sections for higgsino pair production at Run II of the LHC,
i.e.\ $pp$ collisions with 13 TeV centre-of-mass energy, for a range of
particles masses that is relevant for current and future experimental
searches. Similarly in Sec.~\ref{sec:4}, we present the results for a typical
process in the scenario where the neutralinos and charginos have a large
gaugino content. Our conclusions are given in Sec.~\ref{sec:5}.


\section{Theoretical method}
\label{sec:2}

The aim of this work is to study the impact of threshold-resummation
improved PDFs in the calculation of SUSY processes involving neutralinos
and charginos as well as updating NLO+NLL predictions for these cross
sections for ongoing searches at Run II of the LHC, i.e.\ for proton-proton
collisions with a centre-of-mass energy of 13 TeV. For this purpose we adopt
as our PDF baseline the recent global NNPDF3.0 PDF set, which includes data
from ATLAS and CMS analyses on jet, vector-boson and top-quark production
\cite{Ball:2014uwa}. We study the changes that arise when we employ the
threshold-resummation improved PDF set NNPDF30\_nll\_disdytop together with
its fixed-order version NNPDF30\_nlo\_disdytop \cite{Bonvini:2015ira}.
The resummed PDF sets are obtained from a reduced data set involving
deep-inelastic scattering (DIS), Drell-Yan (DY) and top-quark pair
production data, as these are the sole processes for which an analytic
expression of the matrix element at NLO+NLL is available. The reduction of
the data sets adopted in the fit of the threshold-resummation improved PDF
sets leads to a larger PDF error with respect to the global set.

In order to combine the smaller PDF error of the global PDF sets with the
effect of the resummation in the fit of the new PDF sets, we follow the
approach described in Ref.~\cite{Beenakker:2015rna}, where the authors
introduce a factorisation ($K$-factor) method. We define the same $K$-factor
in Eq.\ (\ref{eq:K_factor}),
\begin{equation}
 K = \frac{\sigma({\rm NLO+NLL})_{\rm NLO~global}}{\sigma({\rm NLO})_{\rm NLO~global}} \cdot \frac{\sigma({\rm NLO+NLL})_{\rm NLO+NLL~reduced}}{\sigma({\rm NLO+NLL})_{\rm NLO~reduced}},
 \label{eq:K_factor}
\end{equation}
which can be used to obtain an approximate result for a consistent NLO+NLL
cross section via
\begin{equation}
 \sigma({\rm NLO+NLL})_{\rm NLL+NLO~global} = K \cdot \sigma({\rm NLO})_{\rm NLO~global}.
 \label{eq:K_application}
\end{equation}
This method also allows to rescale the PDF uncertainty extracted
from the global set to a consistent NLO+NLL result. The benefits are twofold:
On the one hand, the relative size of the PDF uncertainties is obtained only from
the global set, which is affected by a smaller PDF error. On the other hand,
we avoid refitting the NNPDF replicas in Mellin space, which typically leads
to some convergence issues \cite{Fuks:2013lya}.

The other main source of theoretical uncertainty is the freedom in the choice
of the unphysical factorisation and renormalisation scales. This uncertainty
will not be rescaled from the NLO results through the $K$-factor. Instead it will be
computed directly. The resummation procedure typically reduces the dependence
of the results on the choice of the scales. Thus we will explicitly estimate
the scale uncertainties on the NLO+NLL result, following the seven-point
method, i.e.\ varying the two scales by relative factors of two, but not four
about the central scale, taken to be the average mass of the two produced SUSY
particles. The total theoretical error will be taken as the sum in
quadrature of the PDF and scale uncertainties.


\section{Higgsino pair production}
\label{sec:3}

Naturalness arguments on the spectrum of SUSY theories require the masses
of higgsinos to be small, i.e. below the TeV scale. In this context one
would also expect a compressed spectrum, where the LSP ($\tilde{\chi}^0_1$),
the lightest chargino ($\tilde{\chi}^\pm_1$) and the next-to-lightest
neutralino ($\tilde{\chi}^0_2)$ are close in mass. Experimental analyses
with the largest sensitivity to this kind of scenario consider three main
processes, which all lead to signatures with soft leptons and moderate
missing transverse momentum in the final state \cite{Sirunyan:2018iwl}.
The first two processes are the associated production of a $\tilde{\chi}^0_2$
and a positively or negatively charged $\tilde{\chi}^\pm_1$, while in the
third process a pair of neutralinos ($\tilde{\chi}^0_2\tilde{\chi}^0_1$)
is produced. The heavier neutralino $\tilde{\chi}^0_2$ and the charginos
$\tilde{\chi}^\pm_1$ will decay to the lighter $\tilde{\chi}^0_1$ through
an off-shell $Z$ or $W^\pm$ boson, respectively. Since the decay products
are expected to be soft because of the compressed spectrum, a jet with
large transverse momentum produced through initial state radiation
can enhance the discriminating power with respect to SM processes
\cite{Sirunyan:2018iwl}.

In the following, we consistently compute the cross sections for the
aforementioned processes at NLO+NLL adopting threshold-improved PDFs and
the $K$-factor method described in Sec.~\ref{sec:2}. The spectra with the
specific characteristics of this scenario have been obtained with the public
code SPheno \cite{Porod:2003um,Porod:2011nf}, following the considerations in
Ref.~\cite{Fuks:2017rio}. In particular, light higgsino-like neutralinos and
charginos $\tilde{\chi}^0_1$, $\tilde{\chi}^\pm_1$ and $\tilde{\chi}^0_2$ of
masses similar to the higgsino mass parameter $\mu$ can be obtained by setting
this parameter to $\mu\leq M_1 = M_2$, i.e.\ below the bino and wino mass
parameters $M_1$ and $M_2$. We choose $\mu$ between 100 GeV and 500 GeV in
order to stay (not too far) above the experimental exclusion limits, while
our choice $M_{1,2}$ = 1 TeV ensures a large higgsino content and mass
splittings of the order of 5 GeV (i.e. $m_{\tilde{\chi}^0_2}-
m_{\tilde{\chi}^\pm_1} \approx m_{\tilde{\chi}^\pm_1} - m_{\tilde{\chi}^0_1}
\approx$ 5 GeV). Our calculations of differential and total cross sections at
NLO+NLL are performed using RESUMMINO \cite{Fuks:2013vua} interfaced with
LHAPDF6 \cite{Buckley:2014ana} for the interpolation of the PDF grids.
The SM parameters have been chosen according to their current PDG values
\cite{Patrignani:2016xqp}, and we fix $\alpha_s(M_Z)=0.118$ with
$\Lambda_{n_f=5}^{\overline{\rm MS}}=0.239$ GeV as appropriate for
NNPDF3.0.

\begin{figure}
 \begin{center}
\includegraphics[width=\textwidth]{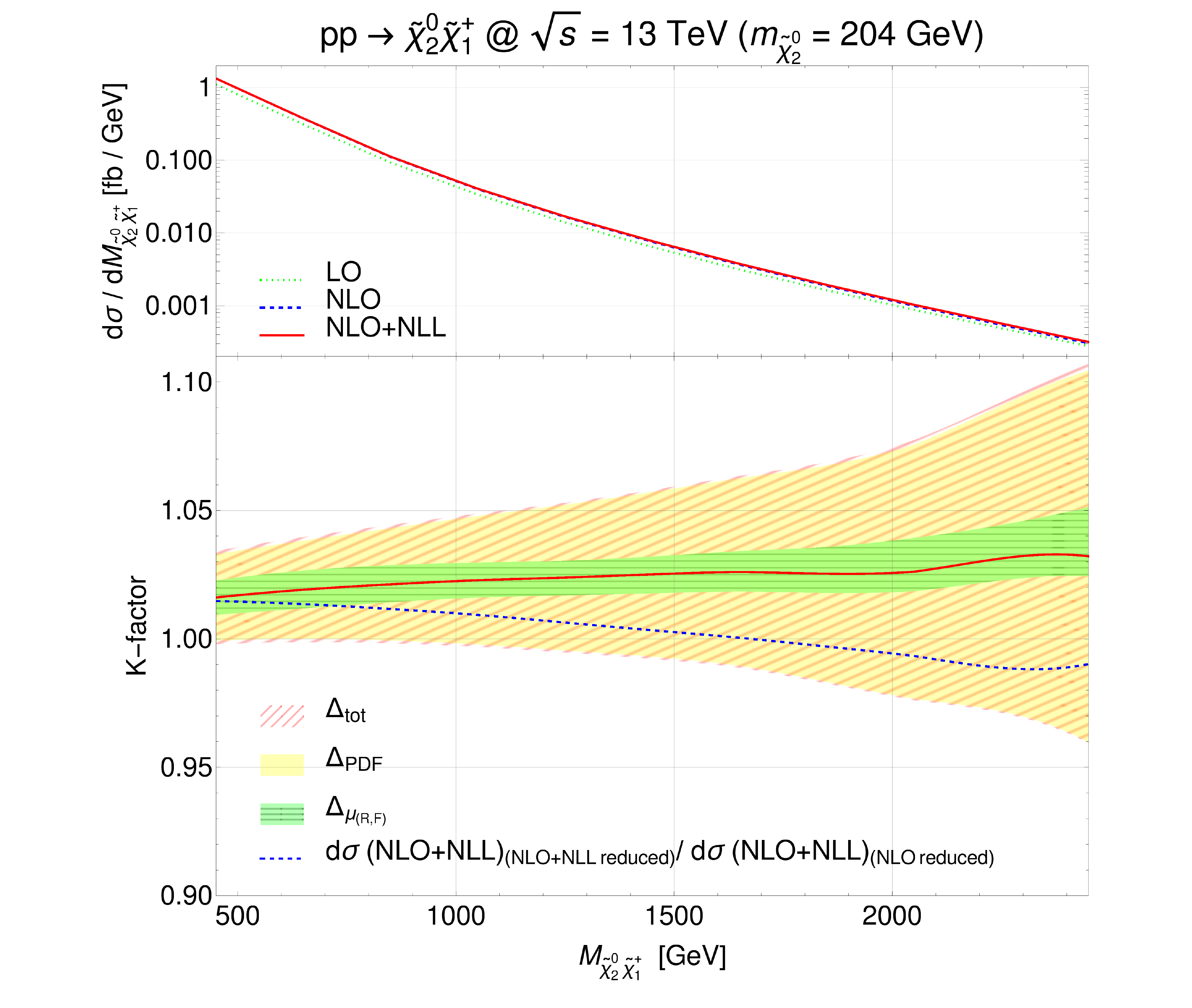}
\caption{Invariant-mass distributions (upper panel) and $K$-factors (lower
 panel) according to Eq.\ (\ref{eq:K_factor}) using the full expression (full
 red) and only its second, PDF-dependent part (dashed blue line) for
 $\tilde{\chi}^0_2\tilde{\chi}^+_1$ production at the LHC with $\sqrt{s} =
 13$ TeV. The higgsino masses are $m_{\tilde{\chi}^0_2}$ = 204 GeV and
 $m_{\tilde{\chi}^+_1}$ = 199 GeV. In the upper panel, the results at LO
 (dotted green), NLO (dashed blue) and NLO+NLL (full red line) have been
 obtained with global NLO PDFs. In the lower panel, the PDF (yellow band) and
 scale (horizontally dashed green band) uncertainties have been computed at NLO
 and NLO+NLL, respectively,
 with global NLO PDFs, then rescaled appropriately and added in quadrature for
 the total theoretical uncertainty (diagonally dashed red band).}
\label{fig:inv_mass_higgsinos}
\end{center}
\end{figure}

We begin with the invariant-mass distribution for the associated production of
a second-lightest neutralino and the lightest positive chargino
($\tilde{\chi}^0_2\tilde{\chi}^+_1$). These differential cross sections at LO,
NLO and NLO+NLL computed with the global NNPDF3.0 PDF set are shown in the
upper panel of Fig.~\ref{fig:inv_mass_higgsinos}.
Comparing the LO and NLO curves we observe an increase of about 20\% in the
low invariant mass region and of about 10\% at higher invariant masses. A
further enhancement of about 1\% and 4\% reflects the effect of the
resummation in the NLO+NLL curve compared with the NLO result in the low and
high invariant-mass regions, respectively. In the lower panel of Fig.\
\ref{fig:inv_mass_higgsinos} we show the $K$-factor that has been defined in
Eq.\ (\ref{eq:K_factor}) (full red) as well as the contribution of its second
term only (dashed blue line), which highlights the effect of the resummation
within the PDF fits. In the low invariant-mass region, the two curves are
similar and predict an increase of the cross section from the global
fixed-order results of about 1.5\%. This can be explained by the observation
that far from threshold resummation effects are small both in the PDFs and even
more so in the partonic matrix elements. In the high invariant-mass region,
the resummation effects in the PDFs reduce the total cross section (dashed
blue), partially compensating those of the partonic matrix elements in the
total result (full red curve). The overall effect is an increase of the
fixed-order result by about 3\%. The error bands represent the PDF (yellow
band) and scale uncertainties (horizontally dashed green band). They have been
calculated and combined into the total theoretical uncertainty (diagonally
dashed red band) following the procedure explained in Sec.~\ref{sec:2}. As one
would expect, the PDF error
grows at high invariant masses due to the larger uncertainties of the PDFs in
the high-$x$ region, whereas the scale error is considerably smaller and stays
relatively constant.

\begin{figure}
\begin{center}
\includegraphics[width=.46\textwidth]{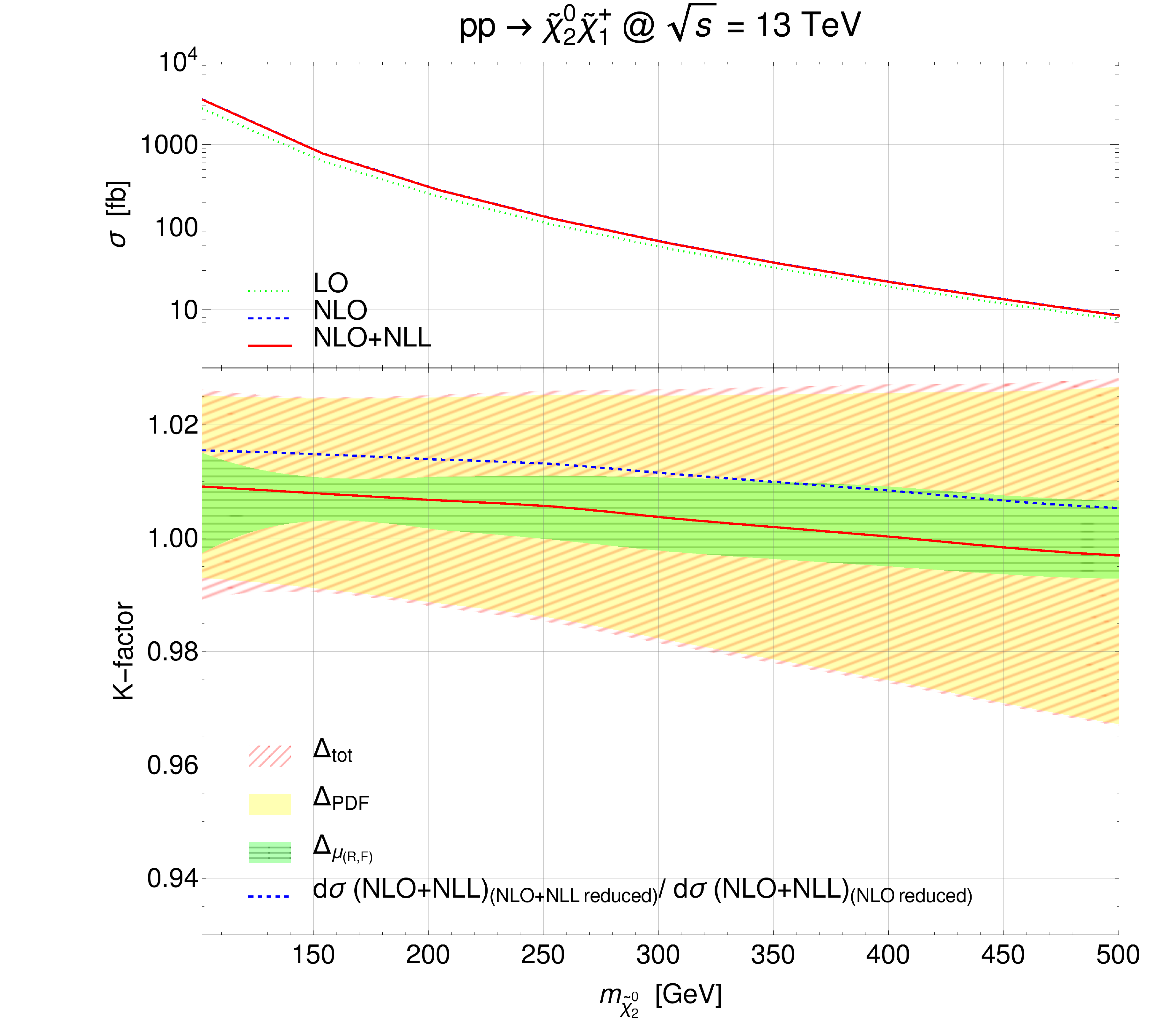}{(a)}
\includegraphics[width=.46\textwidth]{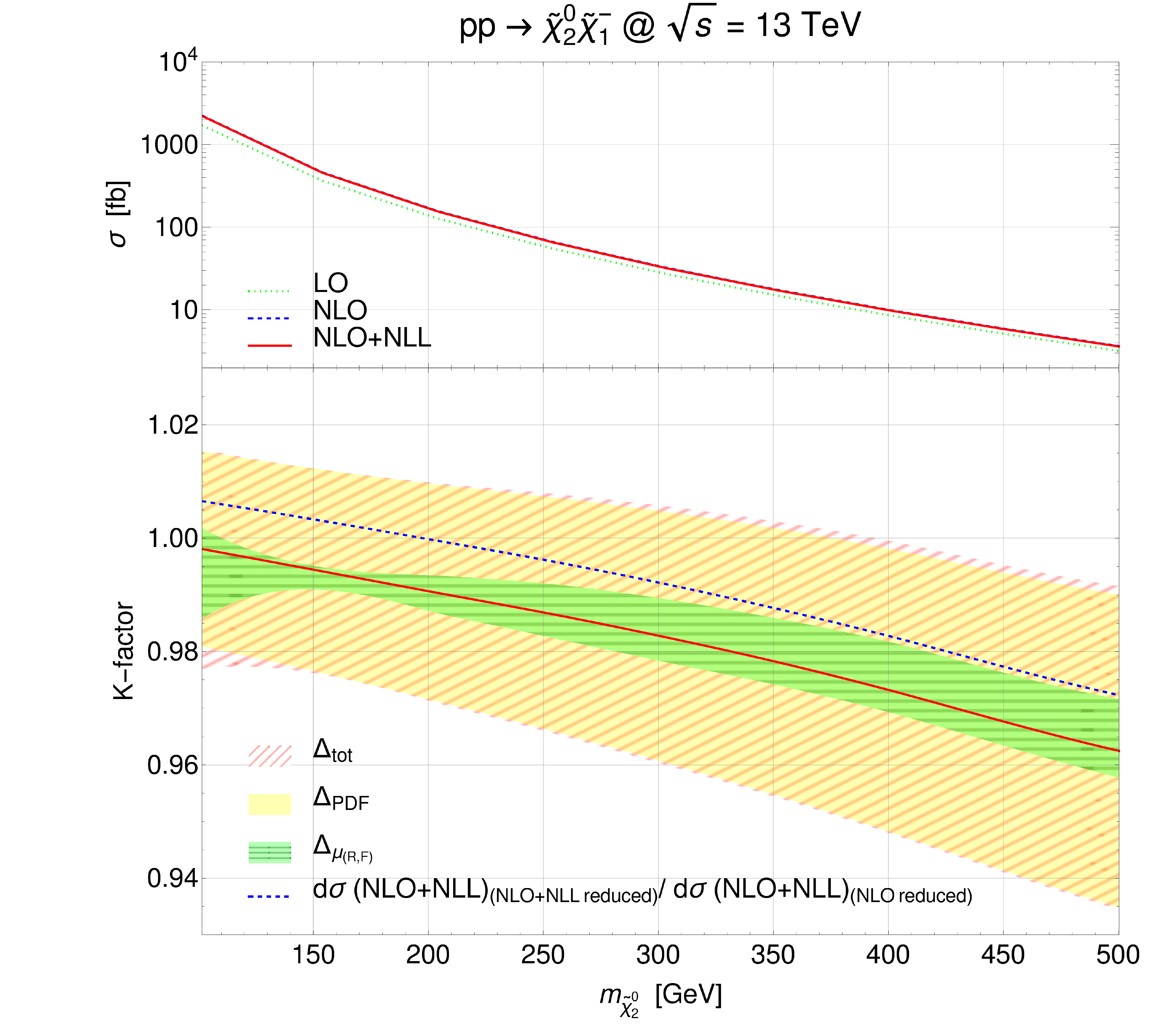}{(b)}
 \caption{Same as Fig.\ \ref{fig:inv_mass_higgsinos}, but for the total cross
 sections of the associated production of second-lightest neutralinos with
 positive (left) and negative (right) charginos as a function of the neutralino
 mass. The chargino is always about 5 GeV lighter.}
\label{fig:mass_higgsinos_pm}
\end{center}
\end{figure}
Following the same procedure, we show in Fig.~\ref{fig:mass_higgsinos_pm}
total cross sections for the associated production of the second-lightest
neutralino and the lightest chargino ($\tilde{\chi}^0_2\tilde{\chi}^\pm_1$)
as function of the mass of the neutralino. The chargino is always about 5 GeV
lighter. We present the results for both positive (left) and negative (right)
chargino production. As expected for a $pp$ collider such as the LHC, the
first of these processes has a larger cross section. Both cross sections are
enhanced by the QCD corrections from LO to NLO by about 30\% for light
higgsino masses and by about 13\% as the particles get heavier. The
resummation in the partonic matrix elements leads instead to a reduction of
the cross section by about 1\% almost independently of the higgsino masses.
In the lower panels we plot the $K$-factors of the two processes. They show
different trends. For positive chargino production (left), the introduction of
the new resummed PDF set (dashed blue line) further increases the cross section
by about 1.5\% at low electroweakinos masses and of about 0.5\% at higher
masses. This effect is partially compensated by the effect of the resummation
in the partonic matrix elements, such that in the end we observe a positive
correction to the fixed order NLO cross section obtained from the global PDF
set of less than 1\% for light higgsino masses and a negative correction
of less than 0.5\% for heavier masses.
The results are different in the case of negative chargino production (right).
The resummation in the PDFs brings a positive correction to the cross section
of about 0.5\% only for very light higgsino masses, and it rapidly turns to
negative values as the higgsinos get heavier and then reaches a negative
correction of about 3\%. This correction gets accentuated by the effect of the
resummation in the partonic matrix elements, so that for light higgsino masses
the  $K$-factor is close to unity, while for heavier masses it gives a
negative correction of about 4\%. In both cases, the total theoretical
uncertainty is dominated by the PDF error, which grows towards larger higgsino
masses, while the scale error stays again relatively constant.
 
\begin{figure}
\begin{center}
\includegraphics[width=\textwidth]{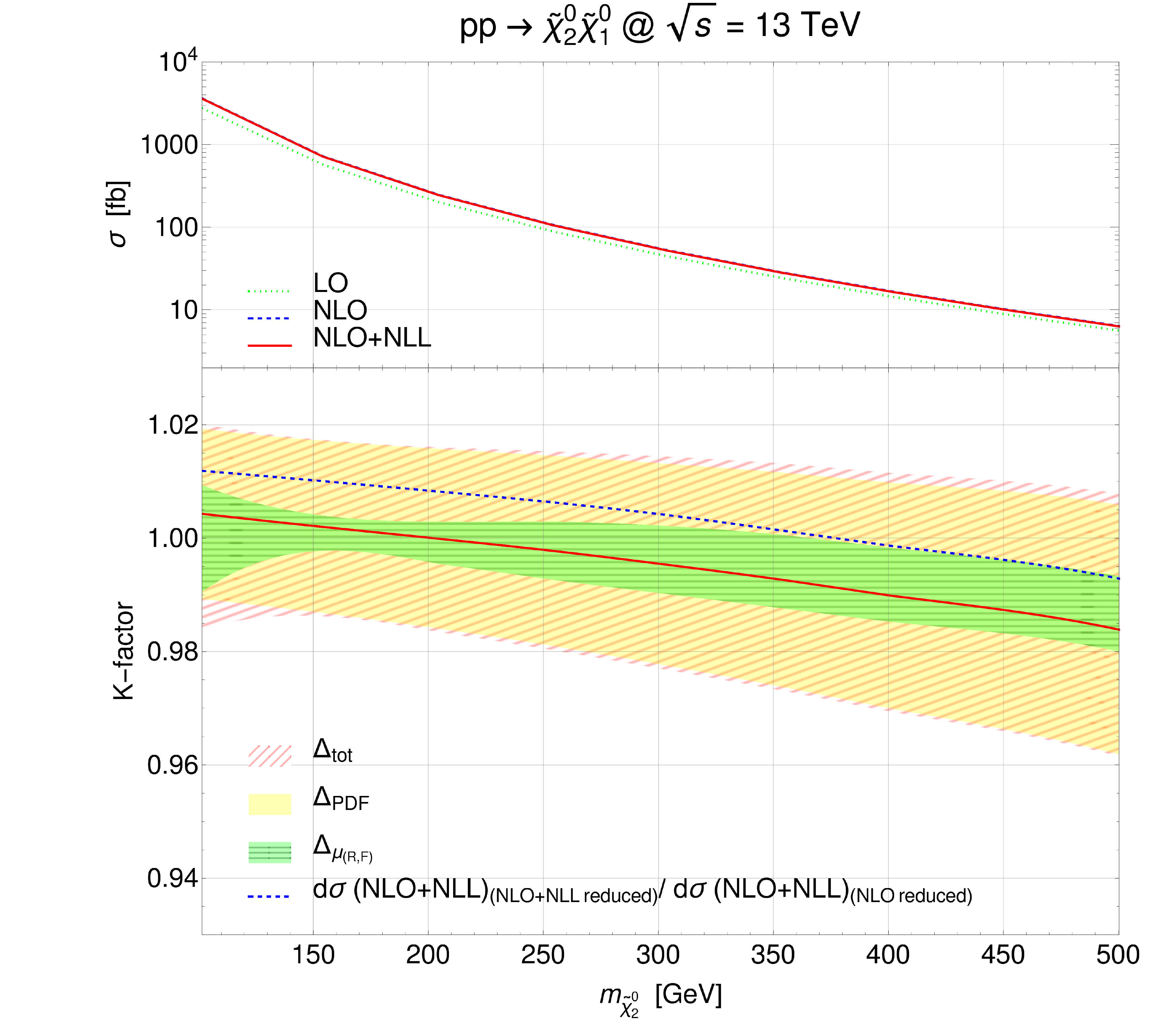}
 \caption{Same as Fig.\ \ref{fig:mass_higgsinos_pm} for total cross sections
 of the associated production of second-lightest neutralinos with the lightest 
 neutralino as a function of the mass of the former. The latter is always about
 10 GeV lighter.}
\label{fig:mass_higgsinos_n}
\end{center}
\end{figure}

In the higgsino scenario, we also consider the process for the associated
production of two neutralinos $\tilde{\chi}^0_2\tilde{\chi}^0_1$. In the
upper panel of Fig.~\ref{fig:mass_higgsinos_n}, we show the cross sections
for this process at LO, NLO and NLO+NLL obtained with the global NNPDF3.0 set.
The effect of QCD corrections is similar to what we have found in the previous
cases. They produce an enhancement of the cross section  from LO to NLO of
about 30\% for light higgsinos and of about 13\% for heavier masses. The
resummation decreases the cross section by about 1\% when going from NLO to
NLO+NLL independently on the higgsino mass. In the lower panel, we can see
that the effect of the threshold-improved PDFs (blue dashed curve) ranges
between $\pm$ 1\% over the neutralino mass range, while the overall effect of
the resummation, both in the PDFs and in the partonic matrix elements (full
red curve), leads to a positive correction of the order of 0.5\% for light
neutralinos and a negative correction of the order of 1.5\% for heavier
masses. The behaviour of the PDF, scale and total theoretical uncertainties
is again similar as before.

We summarise the results of this section by giving explicit values for the
integrated cross sections for the three described processes in Tabs.\
\ref{tab:higgsinos_p}, \ref{tab:higgsinos_m} and \ref{tab:higgsinos_n}.
We include the theoretical uncertainties that have been calculated following
the procedure described in Sec.~\ref{sec:2}. The symmetric PDF errors on the
NLO+NLL cross sections are derived from the application of the $K$-factor,
thus their relative size will be the same for the fixed order and the resummed
results. The asymmetric scale uncertainties instead have been explicitly
estimated in each case through the seven-point method. Following this approach,
the benefits of the resummation on the uncertainty from scale variation are
clearly visible.
\begin{table}
\begin{center}
\begin{tabular}{|c|c||c|c|c|}
  \hline
  $m_{\tilde{\chi}^0_2}$ [GeV] & $m_{\tilde{\chi}^+_1}$ [GeV] & LO (LO global) [fb] & NLO (NLO global) [fb] & NLO+NLL (id.\ global) [fb]\\
  \hline
  $101.5$ & $96.2$ & $2733^{+4.4\%}_{-5.4\%}\pm6.3\%$ & $3545^{+2.1\%}_{-1.7\%}\pm1.6\%$ & $3577^{+0.6\%}_{-1.2\%}\pm1.6\%$ \\
 \hline
  $154.2$ & $148.9$ & $630^{+1.3\%}_{-2.0\%}\pm6.2\%$ & $784^{+1.8\%}_{-1.5\%}\pm1.7\%$ & $790^{+0.3\%}_{-0.5\%}\pm1.7\%$ \\
 \hline
  $204.5$ & $199.0$ & $234^{+0.0\%}_{-0.4\%}\pm6.3\%$ & $284^{+1.9\%}_{-1.3\%}\pm1.8\%$ & $286^{+0.4\%}_{-0.5\%}\pm1.8\%$ \\
 \hline
  $254.3$ & $248.5$ & $107^{+1.4\%}_{-1.7\%}\pm6.3\%$ & $127^{+2.0\%}_{-1.5\%}\pm1.9\%$ & $128^{+0.5\%}_{-0.6\%}\pm1.9\%$ \\
 \hline
  $303.9$ & $297.7$ & $55.3^{+2.5\%}_{-2.6\%}\pm6.4\%$ & $65.0^{+2.1\%}_{-1.7\%}\pm2.2\%$ & $65.2^{+0.7\%}_{-0.6\%}\pm2.2\%$ \\
 \hline
  $353.4$ & $346.7$ & $31.2^{+3.5\%}_{-3.4\%}\pm6.4\%$ & $36.2^{+2.1\%}_{-1.9\%}\pm2.3\%$ & $36.2^{+0.8\%}_{-0.6\%}\pm2.3\%$ \\
 \hline
  $402.6$ & $395.3$ & $18.7^{+4.3\%}_{-4.0\%}\pm6.5\%$ & $21.4^{+2.2\%}_{-2.0\%}\pm2.5\%$ & $21.4^{+0.9\%}_{-0.5\%}\pm2.5\%$ \\
 \hline
  $451.6$ & $443.6$ & $11.7^{+4.9\%}_{-4.6\%}\pm6.6\%$ & $13.3^{+2.2\%}_{-2.1\%}\pm2.8\%$ & $13.3^{+0.9\%}_{-0.5\%}\pm2.8\%$ \\
 \hline
  $500.4$ & $491.5$ & $7.61^{+5.6\%}_{-5.1\%}\pm6.6\%$ & $8.59^{+2.3\%}_{-2.3\%}\pm3.0\%$ & $8.57^{+1.0\%}_{-0.4\%}\pm3.0\%$ \\
 \hline
\end{tabular}
\end{center}
\caption{Total cross sections for associated
 $\tilde{\chi}^0_2\tilde{\chi}^+_1$ production at the LHC with
 $\sqrt{s} = 13$ TeV at LO, NLO and NLO+NLL with consistent PDF choices.
 The central NLO+NLL results are obtained with the $K$-factor method as well
 as the PDF (symmetric) uncertainty at NLO (identical in the last two columns),
 whereas the NLO+NLL (asymmetric) scale uncertainty has been computed
 directly.}
\label{tab:higgsinos_p}
\end{table}
\begin{table}
\begin{center}
\begin{tabular}{|c|c||c|c|c|}
  \hline
  $m_{\tilde{\chi}^0_2}$ [GeV] & $m_{\tilde{\chi}^-_1}$ [GeV] & LO (LO global) [fb] & NLO (NLO global) [fb] & NLO+NLL (id.\ global) [fb]\\
  \hline
  $101.5$ & $96.2$ & $1712^{+4.6\%}_{-5.7\%}\pm6.6\%$ & $2244^{+2.1\%}_{-1.7\%}\pm1.7\%$ & $2240^{+0.4\%}_{-1.2\%}\pm1.7\%$ \\
 \hline
  $154.2$ & $148.9$ & $363^{+1.5\%}_{-2.2\%}\pm6.6\%$ & $457^{+1.7\%}_{-1.4\%}\pm1.8\%$ & $455^{+0.1\%}_{-0.3\%}\pm1.8\%$ \\
 \hline
  $204.5$ & $199.0$ & $127^{+0.0\%}_{-0.4\%}\pm6.7\%$ & $155^{+1.9\%}_{-1.3\%}\pm1.9\%$ & $154^{+0.3\%}_{-0.4\%}\pm1.9\%$ \\
 \hline
  $254.3$ & $248.5$ & $54.8^{+1.4\%}_{-1.7\%}\pm6.7\%$ & $66.0^{+2.0\%}_{-1.5\%}\pm2.1\%$ & $65.1^{+0.5\%}_{-0.4\%}\pm2.1\%$ \\
 \hline
  $303.9$ & $297.7$ & $27.0^{+2.6\%}_{-2.7\%}\pm6.8\%$ & $32.0^{+2.1\%}_{-1.7\%}\pm2.3\%$ & $31.5^{+0.7\%}_{-0.4\%}\pm2.3\%$ \\
 \hline
  $353.4$ & $346.7$ & $14.6^{+3.5\%}_{-3.5\%}\pm6.9\%$ & $17.1^{+2.1\%}_{-1.8\%}\pm2.4\%$ & $16.7^{+0.8\%}_{-0.4\%}\pm2.4\%$ \\
 \hline
  $402.6$ & $395.3$ & $8.38^{+4.3\%}_{-4.2\%}\pm7.1\%$ & $9.72^{+2.2\%}_{-2.0\%}\pm2.6\%$ & $9.45^{+0.9\%}_{-0.4\%}\pm2.6\%$ \\
 \hline
  $451.6$ & $443.6$ & $5.06^{+5.1\%}_{-4.7\%}\pm7.3\%$ & $5.82^{+2.2\%}_{-2.1\%}\pm2.7\%$ & $5.63^{+0.9\%}_{-0.4\%}\pm2.7\%$ \\
 \hline
  $500.4$ & $491.5$ & $3.18^{+5.7\%}_{-5.2\%}\pm7.5\%$ & $3.63^{+2.3\%}_{-2.3\%}\pm2.9\%$ & $3.50^{+0.9\%}_{-0.5\%}\pm2.9\%$ \\
 \hline
\end{tabular}
\end{center}
\caption{Same as Tab.\ \ref{tab:higgsinos_p}, but for associated
 $\tilde{\chi}^0_2\tilde{\chi}^-_1$ production.}
\label{tab:higgsinos_m}
\end{table}
\begin{table}
\begin{center}
\begin{tabular}{|c|c||c|c|c|}
  \hline
  $m_{\tilde{\chi}^0_2}$ [GeV] & $m_{\tilde{\chi}^0_1}$ [GeV] & LO (LO global) [fb] & NLO (NLO global) [fb] & NLO+NLL (id.\ global) [fb]\\
  \hline
  $101.5$ & $92.1$ & $2770^{+5.0\%}_{-6.0\%}\pm6.4\%$ & $3633^{+2.2\%}_{-1.8\%}\pm1.5\%$ & $3649^{+0.5\%}_{-1.4\%}\pm1.5\%$ \\
 \hline
  $154.2$ & $144.4$ & $574^{+1.6\%}_{-2.4\%}\pm6.3\%$ & $721^{+1.8\%}_{-1.5\%}\pm1.5\%$ & $722^{+0.2\%}_{-0.4\%}\pm1.5\%$ \\
 \hline
  $204.5$ & $194.4$ & $202^{+0.0\%}_{-0.3\%}\pm6.3\%$ & $247^{+1.9\%}_{-1.3\%}\pm1.6\%$ & $247^{+0.3\%}_{-0.4\%}\pm1.6\%$ \\
 \hline
  $254.3$ & $243.9$ & $88.6^{+1.3\%}_{-1.6\%}\pm6.3\%$ & $106^{+2.0\%}_{-1.5\%}\pm1.7\%$ & $106^{+0.5\%}_{-0.5\%}\pm1.7\%$ \\
 \hline
  $303.9$ & $293.1$ & $44.5^{+2.4\%}_{-2.6\%}\pm6.3\%$ & $52.6^{+2.0\%}_{-1.7\%}\pm1.8\%$ & $52.3^{+0.7\%}_{-0.5\%}\pm1.8\%$ \\
 \hline
  $353.4$ & $342.0$ & $24.4^{+3.4\%}_{-3.4\%}\pm6.3\%$ & $28.5^{+2.1\%}_{-1.9\%}\pm1.9\%$ & $28.3^{+0.8\%}_{-0.5\%}\pm1.9\%$ \\
 \hline
  $402.6$ & $390.6$ & $14.3^{+4.2\%}_{-4.0\%}\pm6.3\%$ & $16.5^{+2.2\%}_{-2.0\%}\pm2.0\%$ & $16.3^{+0.9\%}_{-0.5\%}\pm2.0\%$ \\
 \hline
  $451.6$ & $438.7$ & $8.77^{+4.9\%}_{-4.6\%}\pm6.3\%$ & $10.04^{+2.2\%}_{-2.1\%}\pm2.1\%$ & $9.91^{+0.9\%}_{-0.4\%}\pm2.1\%$ \\
 \hline
  $500.4$ & $486.2$ & $5.60^{+5.6\%}_{-5.1\%}\pm6.4\%$ & $6.36^{+2.3\%}_{-2.3\%}\pm2.2\%$ & $6.26^{+1.0\%}_{-0.4\%}\pm2.2\%$ \\
 \hline
\end{tabular}
\end{center}
\caption{Same as Tab.\ \ref{tab:higgsinos_p}, but for associated
 $\tilde{\chi}^0_2\tilde{\chi}^0_1$ production.}
\label{tab:higgsinos_n}
\end{table}


\section{Gaugino pair production}
\label{sec:4}

We now turn to the case where the produced neutralinos and charginos have
a large gaugino component. The next-to-lightest neutralino
$\tilde{\chi}^0_2$ and the charginos $\tilde{\chi}^\pm_1$ will be
considered as wino-like and almost degenerate with a mass above 760 GeV
to satisfy experimental constraints, while the LSP $\tilde{\chi}^0_1$ is
assumed to be bino-like and light. In this scenario, large production
cross sections of $\tilde{\chi}^0_2\tilde{\chi}^\pm_1$ and short
decay chains are expected. Assuming an intermediate and equal mass for
left-handed staus and tau sneutrinos, the winos will decay through these
states into the LSP, taus and tau neutrinos, leading to interesting
collider signatures \cite{Aaboud:2017nhr}. This particular spectrum of
particle masses can be achieved within the phenomenological MSSM (pMSSM)
framework. It is of particular interest, since the coannihilation of
light staus with the LSP can generate a dark matter relic density in
accordance with the observations.

A spectrum with these features is obtained using the public code SPheno
\cite{Porod:2003um,Porod:2011nf} by setting a small value for the bino
mass parameter $M_1$, while the wino mass parameter $M_2$ is chosen above
the ATLAS exclusion limits. The large gaugino content can be achieved by
setting a large value for the $\mu$ parameter ($\mu \gg M_2$). With this
configuration, only a very small splitting between the masses of the
neutralino $\tilde{\chi}^0_2$ and the charginos $\tilde{\chi}^\pm_1$ is
generated.

We now study the effect of the inclusion of threshold-resummed PDFs in a
consistent calculation of the cross sections at NLO+NLL. We first consider
a specific configuration of the masses and study the invariant-mass
distribution for $\tilde{\chi}^0_2\tilde{\chi}^+_1$ associated production.
\begin{figure}
\begin{center}
\includegraphics[width=\textwidth]{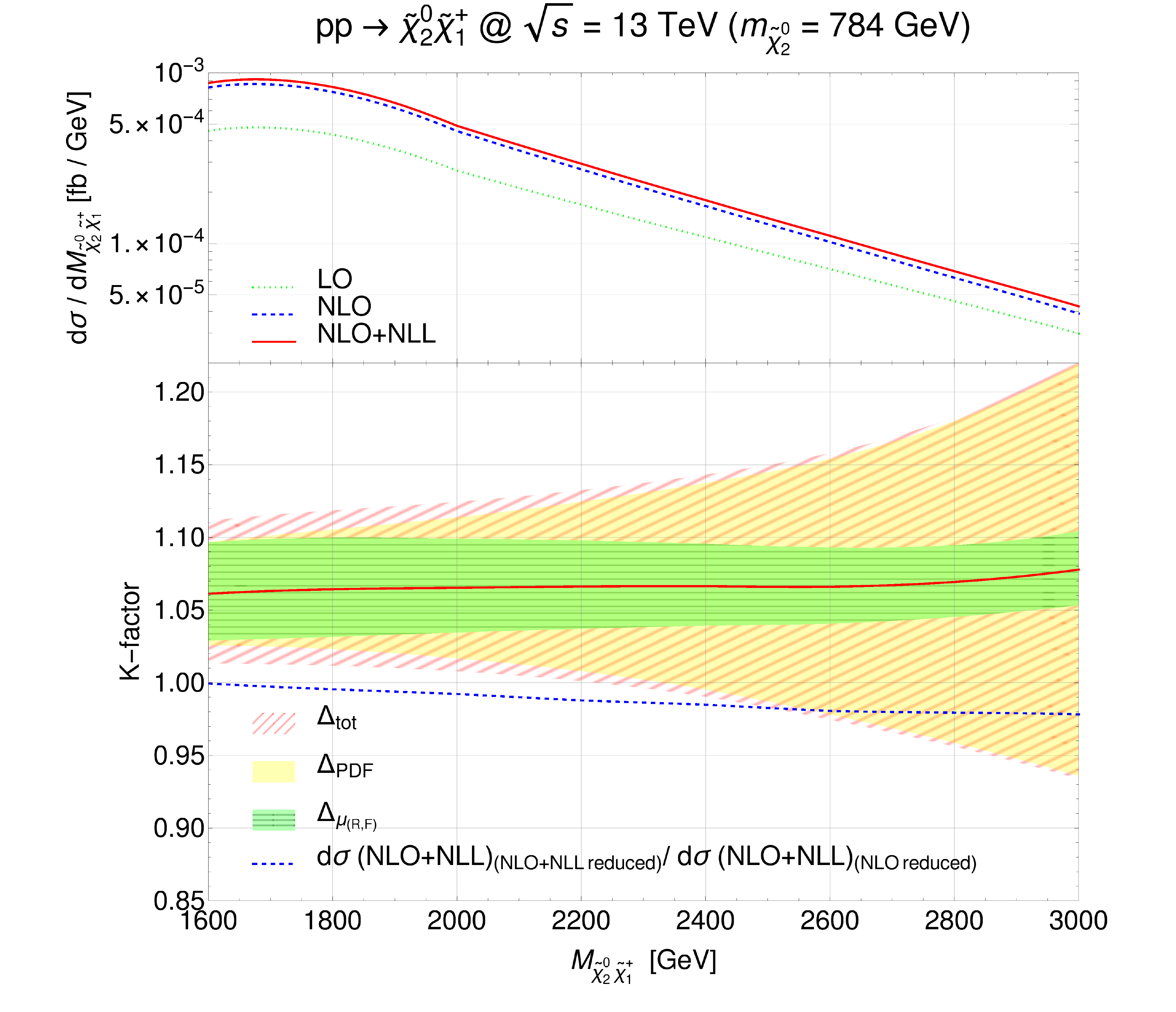}
\caption{Invariant-mass distributions (upper panel) and $K$-factors (lower
 panel) according to Eq.\ (\ref{eq:K_factor}) using the full expression (full
 red) and only its second, PDF-dependent part (dashed blue line) for 
 $\tilde{\chi}^0_2\tilde{\chi}^+_1$ associated production at the LHC with
 $\sqrt{s} = 13$ TeV. The wino masses are $m_{\tilde{\chi}^0_2} \simeq
 m_{\tilde{\chi}^+_1}$ = 784 GeV. In the upper panel, the results at LO
 (dotted green), NLO (dashed blue) and NLO+NLL (full red line) have been
 obtained with global NLO PDFs. In the lower panel, the PDF (yellow band) and
 scale (horizontally dashed green band) uncertainties have been computed at NLO
 and NLO+NLL, respectively, with global NLO PDFs, then rescaled appropriately
 and added in quadrature for the total theoretical uncertainty (diagonally
 dashed red band).}
\label{fig:inv_mass_gauginos}
\end{center}
\end{figure}
In Fig.~\ref{fig:inv_mass_gauginos}, the upper panel shows the invariant
mass distributions calculated at LO, NLO and NLO+NLL with the global
NNPDF3.0 PDF set. Here, QCD corrections have a large impact on the cross
section, which increases from LO to NLO by about 80\% and 30\% at low and
high invariant masses, respectively. The resummation further increases the
cross section by about 6\% in the low invariant-mass region and by about
10\% for higher invariant masses. Resummation effects in the PDFs only
(dashed blue line) are small as they remain below 2.5\% over the whole
invariant-mass region considered here. In the same interval, resummation
in the partonic matrix elements gives a large contribution and produces
an enhancement of the cross section between 6\% and 8\%. The reason is
that the winos in this section are considerably heavier than the
higgsinos in the preceding section, so that we are closer to threshold
and resummation effects are more important.

\begin{figure}
\begin{center}
\includegraphics[width=\textwidth]{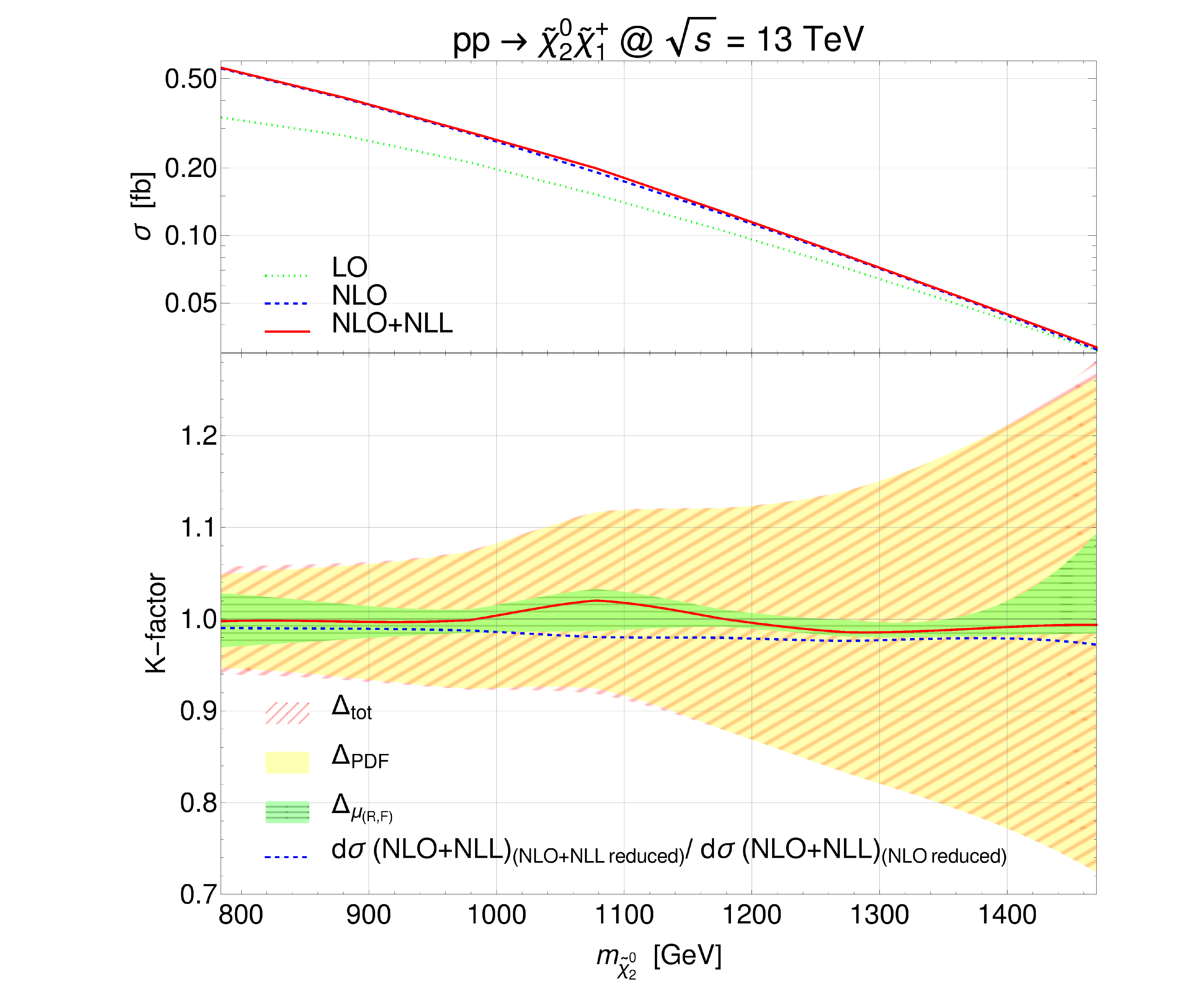}
 \caption{Same as Fig.\ \ref{fig:mass_higgsinos_pm}, but for the total
 cross sections of the associated production of second-lightest neutralinos
 with positive charginos as a function of their mass.}
\label{fig:mass_gauginos}
\end{center}
\end{figure}
We repeat the analysis to study the effects of resummed PDFs on the integrated
cross section. The results are shown in Fig.~\ref{fig:mass_gauginos}. In
the upper plot we can observe the effects of the QCD corrections, which are
large for light winos and enhance the cross section from LO to NLO by about
65\%. The enhancement then rapdily decreases for heavier masses. The effect
of the resummation further increases the cross section from NLO to NLO+NLL by
about 2\% to 4\% in the wino mass range considered here. The effect of the
resummation in the PDFs (dashed blue line) is small in this scenario, and the
combined effect with the resummation in the partonic matrix elements is
dominated by the latter and again it brings a positive contribution to the
cross section between 2\% and 4\% in the whole considered wino mass range.

We conclude this section by giving explicit results for the cross sections
at LO, NLO and NLO+NLL in Tab.~\ref{tab:gauginos}. They have been consistently
calculated through the $K$-factor method in this scenario of neutralinos and
gauginos with large gaugino content.
\begin{table}
\begin{center}
\begin{tabular}{|c||c|c|c|}
  \hline
  $m_{\tilde{\chi}^0_2}$ [GeV] & LO (LO global) [fb] & NLO (NLO global) [fb] & NLO+NLL (id.\ global) [fb]\\
  \hline
  $784.2$ & $0.336^{+8.6\%}_{-7.5\%}\pm7.6\%$ & $0.556^{+8.6\%}_{-7.3\%}\pm5.1\%$ & $0.554^{+3.1\%}_{-2.8\%}\pm5.1\%$ \\
 \hline
  $881.8$ & $0.278^{+9.2\%}_{-8.0\%}\pm8.0\%$ & $0.407^{+7.6\%}_{-6.7\%}\pm6.1\%$ & $0.406^{+1.9\%}_{-2.0\%}\pm6.1\%$ \\
 \hline
  $979.8$ & $0.211^{+9.9\%}_{-8.5\%}\pm8.7\%$ & $0.284^{+7.0\%}_{-6.3\%}\pm7.5\%$ & $0.284^{+1.2\%}_{-1.4\%}\pm7.5\%$ \\
 \hline
  $1078$ & $0.152^{+10.6\%}_{-9.1\%}\pm9.6\%$ & $0.192^{+6.6\%}_{-6.1\%}\pm9.4\%$ & $0.196^{+1.2\%}_{-3.3\%}\pm9.4\%$ \\
 \hline
  $1176$ & $0.106^{+11.3\%}_{-9.6\%}\pm10.8\%$ & $0.126^{+6.4\%}_{-6.1\%}\pm12.0\%$ & $0.126^{+1.0\%}_{-0.9\%}\pm12.0\%$ \\
 \hline
  $1274$ & $0.071^{+12.0\%}_{-10.1\%}\pm12.4\%$ & $0.080^{+6.3\%}_{-6.1\%}\pm15.6\%$ & $0.079^{+1.4\%}_{-0.5\%}\pm15.6\%$ \\
 \hline
  $1372$ & $0.047^{+12.6\%}_{-10.6\%}\pm14.5\%$ & $0.050^{+6.3\%}_{-6.1\%}\pm20.5\%$ & $0.050^{+1.7\%}_{-0.8\%}\pm20.5\%$ \\
 \hline
  $1470$ & $0.031^{+13.2\%}_{-11.0\%}\pm17.2\%$ & $0.031^{+6.2\%}_{-6.2\%}\pm27.2\%$ & $0.031^{+10.1\%}_{-1.0\%}\pm27.2\%$ \\
 \hline
\end{tabular}
\end{center}
\caption{Same as Tab.\ \ref{tab:higgsinos_p}, but for winos instead of
 higgsinos.}
\label{tab:gauginos}
\end{table}


\section{Conclusion}
\label{sec:5}

In this paper, we have studied the effects of the introduction of
threshold-resummation improved PDF sets in consistent NLO+NLL calculations
for the associated production of neutralinos and charginos at Run II of the
LHC. In particular, we computed LO, NLO and NLO+NLL cross sections for
various processes relevant for current and future experimental searches.
The SUSY particles were considered in two different mass ranges and in
scenarios where they featured either a large higgsino or a large gaugino
content.

The impact of the resummation within the PDFs has been parametrised through
a factorisation method employed previously for squarks, gluinos and sleptons.
Using ratios of resummed and fixed-order cross sections in a specific
$K$-factor, it is easily possible to rescale the fixed-order NLO results to
consistent NLO+NLL calculations, which include the effects of the resummation
in the partonic matrix elements as well as in the PDF fits. As expected, the
use of threshold-resummation improved PDFs partially compensates the
enhancement of the cross sections due to the resummation in the partonic
matrix elements.

Scale uncertainties and PDF error bands were given along with the central
values for the $K$-factors. The latter were extracted from the global
NNPDF3.0 set in order to minimise the impact of the reduction of the data
set in the fit of the threshold-resummation improved PDFs. Through this
method, also the troublesome refitting in Mellin space of the NNPDF replicas
of the PDF sets from reduced data sets was bypassed. The effects of the
variation of factorisation and renormalisation scales were determined
explicitly with the corresponding cross sections in order to preserve the
benefits of the resummation in terms of the reduction of scale uncertainties.

To conclude, the presented results allow to further improve the reliability of
the theoretical calculations for the interpretation of the experimental data
during the ongoing LHC Run-II programme.


\acknowledgments

This work has been supported by the BMBF under contract
05H15PMCCA and the DFG through the Research Training Network 2149
``Strong and weak interactions - from hadrons to dark matter''.


\bibliographystyle{apsrev4-1}
\bibliography{bib}


\end{document}